\documentclass[sn-basic]{sn-jnl}

\usepackage{graphicx}%
\usepackage{multirow}%
\usepackage{amsmath,amssymb,amsfonts}%
\usepackage{amsthm}%
\usepackage{mathrsfs}%
\usepackage[title]{appendix}%
\usepackage{xcolor}%
\usepackage{textcomp}%
\usepackage{manyfoot}%
\usepackage{booktabs}%
\usepackage{algorithm}%
\usepackage{algorithmicx}%
\usepackage{algpseudocode}%
\usepackage{listings}%
\usepackage{tikz}
\usepackage{floatflt}
\usepackage[table]{xcolor}
\usepackage{array}
\usepackage{subcaption}

\theoremstyle{thmstyleone}%
\newtheorem{theorem}{Theorem}
\newtheorem{proposition}[theorem]{Proposition}%

\theoremstyle{thmstyletwo}%
\newtheorem{example}{Example}%
\newtheorem{remark}{Remark}%

\theoremstyle{thmstylethree}%
\newtheorem{definition}{Definition}%

\begin{document}

\title[A Visual Discrete Event-based Simulator for Protection of Plants against Herbivores Employed as Computational Optimization Game]{A Visual Discrete Event-based Simulator for Protection of Plants against Herbivores Employed as Computational Optimization Game}


\author[1]{\fnm{Lucas} \sur{Dietrich}}\email{lucas.dietrich@tuhh.de}

\author[2]{\fnm{Benjamin} \sur{F\"orster}}\email{\{bfoerster,langendoerfer\}@ihp-microelectronics.com}

\author[2]{\fnm{Peter} \sur{Langend\"orfer}}

\author*[1]{\fnm{Thomas} \sur{Hinze}}\email{thomas.hinze@uni-jena.de}

\affil*[1]{\orgdiv{Department of Bioinformatics}, \orgname{Friedrich Schiller University}, \orgaddress{\street{Ernst-Abbe-Platz 2}, \city{Jena}, \postcode{07743}, \country{Germany}}}

\affil[2]{\orgdiv{IHP Leibniz Institut f\"ur innovative Mikroelektronik}, \orgname{Brandenburg University of Technology Cottbus-Senftenberg}, \orgaddress{\street{Im Technologiepark 25}, \city{Frankfurt an der Oder}, \postcode{15236}, \country{Germany}}}

\abstract{Plants come with sophisticated strategies to survive within a highly competing environment. In addition, they need to resist frequent attacks from a variety of herbivores acting alone, in small groups, or in swarms. Since the amount of energy a plant might invest in defense and reproduction is limited, a complex optimization problem emerges. In a shared habitat, plants fight herbivores by shape and camouflage, by the release of specific toxins, or by attracting predators of herbivores. Furthermore, plants alert their surrounding field by signaling substances in the event of an assault. Transported by air or through a network of roots, signaling substances reach neighbors to trigger their defense. The offsprings of a plant commonly grow within a certain distance to benefit from symbiotic protection. We introduce a grid-based visual simulation software for detailed configuration and subsequent processing of the behavior of the resulting system in time and space. In terms of solution to a computational optimization problem inspired by nature, settings with low energy need and long life able to cope with different patterns of attack can be figured out and analyzed. Applications include novel techniques for efficient construction and secure operation of sensor networks.}

\keywords{plant defense, event-based simulation, topological optimization}



\maketitle

\section{Introduction}\label{sec1}

Plants created a fascinating variety of mechanisms to protect themselves and their conspecifics or symbiotic partners against manifold herbivorous insects, predatory animals, and microbial pathogens \cite{bhatla18}. Since most plants reside in a fixed position on earth with roots directed toward the ground and aerial leaves harvesting sunlight, they cannot move as a whole in order to escape. Typically, their offsprings settle within the closest environment, but keeping a certain distance to each other. It should be noted that plants of the same species tend to form networks of individuals spread over a region. They share the ground with other plants, trees, and mushrooms in a mixed neighborhood. The resulting habitat acts in a highly competitive manner with symbiotic elements emerged to cope with lack of resources, nutrients, and individual space \cite{zhang21}.

The optimal placement of their offsprings in the surrounding landscape turns out to be merely one challenge for a plant or a family of plants that aims to survive and reproduce. A much greater threat is caused by a variety of \textit{herbivores}. They preferably feed on leaves but also on flowers, or other components of plants for nutrition. Herbivores exist in many forms. Some of them mostly act alone like deer or giraffes. Others favor small groups such as wild pigs or bunnies. In particular, swarms consisting of hundreds or thousands of individuals can be destructive in a disastrous way. Insects, worms, ants, and other articulate animals have the ability to damage wide areas in short time up to the extinction of plant species.

In general, plants have been equipped with a plethora of different and fine-tuned methods and techniques for protection against herbivores \cite{burow17}. The genomes found in the flora belong to the largest among all life forms. Although there is no central nervous system or brain, a plant typically realizes a feed attack since the destroyed cell structures at the edge of a bite mark become contaminated. The presence of foreign molecules left by the herbivorous animal initiates a biochemical signaling cascade which in turn can be kept locally or - in case of a larger attack - incorporates most parts of the plant to get activated. The amount, spatial distribution, and combination of foreign molecules reveal information about the enemies and identify the type of attacker along with the intensity and directions of feed. Now, signaled cells around the locations of damage can start a specific defense.

Defense and deterrence consume \textit{energy}, a very limited resource. Young and small plants with low \textit{leaf area} in total might harvest much less energy from sunlight than full-grown trees with approximately $1,200\mathrm{m}^{2}$ on average up to a maximum of around $15,000\mathrm{m}^{2}$ \cite{carreiro08}. Root ends mainly contribute water and nutrients, but rarely exploitable energy. Metabolism and life support require an individual threshold of permanent energy need for existence. Energy resources beyond the threshold are available for reproduction and protection.

The behavior of plants in terms of energy use reflects the process of solving a complex \textit{optimization problem} about how to survive and reproduce with a minimal amount of energy fending off all relevant herbivores. In some cases, it seems to be advantageous to spend energy for the permanent development of prickles, thorns, or leaves with saw-toothed shape \cite{chakraborty23}. In this way, single individuals of larger animals might lose their willingness to feed. The same holds for the release of bitter substances which in turn need to be produced by the plant consuming energy. Ideally, the benefit consists in the generation of a memory effect, and herbivorous animals remember their experience and avoid the plant for long time. In contrast, this strategy fails at swarm attacks. Here, the plant must invest much more energy in the rapid production of effective specific toxins or drugs. While toxins aim at a deadly effect that kills the individual attacker, drugs diminish the alertness and awareness of attackers in combination with attracting their enemies. Moreover, families of plants can alert each other. If one plant is attacked, it sends signaling substances to other members of its family. This can happen by spreading over air, but also through a mycorrhiza, an underground symbiotic association between roots and a fungus. The alerted plants then start to prepare for defense on their own. So, a swarm can be eliminated in a short time, but with a high use of energy. Especially families of plants are capable of regeneration after some members have died off. The clue might be found in an optimal spatial distribution of the plants in the landscape. The resulting topology of the vegetation cover in total gives a solution to the optimization problem mentioned above. In addition, the response scheme of individual plants against herbivores marks a part of the optimal solution.

What stands out is the idea of a software simulator that enables a fine-grained configuration of a scenario consisting of all players like families of plants, clusters of herbivores, and a pool of signaling molecules, bitter substances, and toxins at an abstract level captured by appropriate parameters and attributes. The initial distribution of plants and herbivores is done on a grid in a freely configurable manner. After the whole setting had been specified, the simulation runs. Its progression is controlled by discrete time steps and events. Typical events are for instance the update of the energy level of an individual plant, the elimination of a plant whose energy level fell under the predefined threshold for survival, placement of the offsprings of a plant, loss of herbivore individuals, motion of a cluster of herbivores on the grid, and many others. Performing an event might imply the generation of additional events at later time points. In conjunction with simulation, the behavior of the entire system becomes visible and can be analyzed by the software on the fly. This feature allows for comparison of different initial settings of systems with respect to energy need, reproduction rate, and overall fitness. In the next stage, an optimization towards robust and persistent families of plants with low energy need is feasible. To the best of our knowledge, this is the first attempt to provide such a tool adopting natural computation in a moderately abstract processing scheme for the protection of plants against herbivores. Our software named \textit{PlantProtectionSim} is freely available from Github\footnote{\url{https://github.com/lucas-diet/PlantProtectionSim}}.

The security of \textit{wireless sensor networks} (WSNs) can be informed by strategies observed in plant defenses, as both systems are resource-constrained and must operate reliably in hostile environments \cite{foerster24}. In such a situation, hundreds to thousands of sensor nodes are deployed to carry out monitoring or controlling tasks like wildfire detection, precision agriculture (for example, irrigation control), or infrastructure monitoring (for example, bridges). Like plants, WSNs cannot maintain absolute protection and must balance the energy expenditure of defense with long-term operational needs. Plants combine constitutive and inducible mechanisms and even signal neighboring organisms when under attack \cite{foerster23}. This biological analogy motivates the design of distributed and collaborative security schemes for WSNs, aiming to broaden threat coverage while limiting the burden on individual nodes.

The paper is structured as follows. Section \ref{sec2} sketches the biological background and reflects related work. In Sections \ref{sec3} and \ref{sec4}, we describe the underlying model and its implementation for the simulator software, while Section \ref{sec5} is dedicated to a user manual in brief. A case study will be addressed in Section \ref{sec6}. Eventually, Section \ref{sec7} discusses the results, draws conclusions, and sheds light on future work.

\section{Biological Background and Related Work}\label{sec2}

Plant protection strategies against potential attackers can be classified into \textit{constitutive}, \textit{induced}, and \textit{activated} defense mechanisms from a systems biology perspective \cite{blande22,vos16}. All of them come with specific advantages worth to be generalized towards technical counterparts.

The simplest way for small plants to survive consists of getting \textit{invisible} if necessary. Grass follows this strategy, for example. It can grow rather fast, stores nutrients in its roots, and it is able to survive for weeks without any contact with sunlight. After a predator has fed off the halms, the grass regenerates beneath the ground. This behavioral strategy does not require effort or resources for a dedicated defense or alert. For the long-term survival of grass in the savanna, this seems to be successful, but grass is restricted in its size and in its ability to persist in the relative darkness of forests. When plants started to compete with each other for the maximal portion of sunlight, they needed to become larger and more durable, which implies more sophisticated strategies against attackers. If a plant is under potential threat through an attacker, it makes use to evaluate whether a countermeasure will be expedient, since a reaction also causes costs and stress to it \cite{sopory19}.

In case a plant does not actively react, there is a \textit{constitutive} defense mechanism in place that protects the plant against a multitude of potential threats without additional costs \cite{vos13}. Examples of constitutive defense mechanisms are spikes or tooth-shaped outer forms of leaves grown by nature in the plant. Complementary, some plants continuously accumulate poisonous chemicals in their leaves or stipes. Here, chemicals commonly arise as side products of the underlying metabolism. A typical example is tobacco, which produces nicotine in this way, a mild poison that affects the nervous system of herbivores. A brindle coloring of the leaves helps a couple of plants in camouflage. All constitutive defense mechanisms have in common that no energy-consuming signaling is necessary inside and among plants. Generally, the constitutive defense mechanism makes herbivores avoid the corresponding plant and exemplars of the same species for a certain time. Rarely does this strategy lead to the death of herbivores. The additional energy need for the plant remains low, but its genome requires extension.

A more complex defense pattern is defined by \textit{induced} security mechanisms whose main feature lies in a high precision against the specific attacker in combination with an outstanding flexibility to cope with different types of enemies \cite{pieterse14}. The behavioral pattern for defense gets induced upon an ongoing attack or in case an attack is signaled by other surrounding plants of the same species. Plants react in a purposive manner. The resulting cascade of biochemical response comprises successive steps of action and interaction. Typically, the processing scheme starts with a form of \textit{signaling} \cite{baluska22}. Herbivores destroy the inner structure of leaves or stipes, which in turn implies one or more wounds. In addition, they leave indications about their identity by contamination of the wounds. The first stage of signaling releases messenger substances that circulate within the affected plant. The presence of those messengers initiates a dedicated biochemical production of response molecules. They in turn initiate a \textit{chemical defense} by opening reservoirs of highly concentrated poisons or chemicals that cause a bad taste directly transported to the wounds \cite{poppenga10}. Alternatively, direct defense can attract herbivore enemies using pheromones or attractants. A third type of defense revokes the chlorophyll of their leaves, which stops photosynthesis, but also eliminates nutrients from their leaves. This comes along with a temporary change in the colors of leaves subsumed by the term \textit{aposematic signaling} \cite{lev16}. Furthermore, the plant starts to inform its neighbors about the ongoing attack. The corresponding second stage of signaling acts in terms of an alert.

For this purpose, the corresponding signals exhibit a biochemical nature \cite{gatehouse02}. Although some gaseous chemical signals such as ethylene can be communicated over air, a variety of protein-based signals are exclusively communicated via roots of plants and fungi and often through the symbiotic mycorrhiza \cite{bhattacharyya15}. Typically, these messaging substances make use of electrical forces for transport. An alert signaled from plants in its neighborhood has the same effect as a direct attack. After a plant has received an alert and the concentration of involved signaling molecules exceeds a certain threshold, the chemical defense or aposematic signaling is set into operation. Induced security mechanisms run for a certain amount of time after an attack occurs or has been signaled. Then, they stop and turn into regeneration of the plant.

Induced defense mechanisms turn out to be precise and productive \cite{chen08}. Herbivores are usually devitalized or even eliminated. Some animals learn to circumvent the effect of toxins, especially in the case of signaling over air. They simply feed against the wind direction. The plant must have resources to exploit a large amount of energy in a short time to produce all the required substances by means of the corresponding genes.

\textit{Activated} defense mechanisms complete the variety of strategies found in plants \cite{bruce15}. Once triggered, those mechanisms remain permanently active until the plant dies. An example is an unstable toxin generation in varying leaves with random intensity over time. After the production of those toxins has been triggered, the plant will produce them sporadically in irregular periodicity and at different locations in the plant. As soon as such a defense mechanism is activated, it is no longer reversible and consumes energy continuously. For scenarios in which waves of attacks appear in an iterative manner, activated defense mechanisms save time and enable a much faster response in spite of the high energy costs.

Each of the chosen concepts has an impact on the fitness of the plant. Therefore, it limits the frequency and complexity of the reactions that a plant can impose. The goals of different defense mechanisms are either to stop an attack or limit the caused damage to the current target or future targets represented by nearby plants. 

The chemical signaling process is the primary process for informing other plants about ongoing attacks in their environment. The most common inter-plant signaling concept with the largest reach is the chemical signaling process \cite{sun21}. Plants produce a multitude of messenger molecules. Those can spread through air to warn neighboring plants about threat scenarios \cite{farmer01}. Different types of molecules impose different fitness costs to the plant. In addition, they have different molecular mass and therefore the distances vary, and these molecules can bridge and inform other plants about different threat scenarios. Moreover, it has to be noticed that for particular types of chemical signals, there is a corresponding specific subset of plants that understand them. For this reason, plants might signal each other hierarchically. 

Hence, a plant produces molecule $A$ as signals which is understood by a subset of surrounding plants, which in turn start producing the chemical signal $B$ alerting a different subset of plants. The result is a dynamic mixture of chemicals in the air that always surround those plants. Each plant has a concept of  biological signal processing implemented on a cellular level \cite{aljbory18}. With that, the cells are able to distinguish and evaluate the different signals. Furthermore, plants decide about their behavior pattern based on activation functions and biases about whether and how to react to potential threats \cite{burow17}.

The long-term process of mutual interaction between plants and herbivores results in complex behavioral patterns along with the optimal spatial distribution of plants on the ground and optimal exploitation by herbivores, respectively. Having a plethora of biological data at hand, a deeper understanding of the interplay of defense mechanisms acting as a whole becomes feasible. To do so, the first step is to develop simulation software that integrates all relevant biological aspects in an appropriate level of abstraction.

\section{Modeling and Implementation of the Simulator Software}\label{sec3}

The user can configure scenarios within the simulation system in a relatively flexible manner, combining plants, their spatial distribution and interactions, as well as available protective mechanisms, together with herbivores and their respective abilities and traits. In certain respects, the simulation system resembles a computer game in which a set of actors is first defined and configured, after which these actors encounter one another and interact dynamically. We arrange the software to mimic a visual discrete event-based simulator due to its ability to cope with rapid local changes.

\subsection{Configuration of Plants}

As a representation of the flat landscape, a rectangular grid of up to $80$ by $80$ unit squares is employed, hereafter referred to as \textit{grid cells}. Each cell may be occupied by no more than a single plant. The user specifies the number of \textit{plant species} to be distinguished (at least one, at most sixteen). Each species is assigned a distinct color. A plant is placed on a cell by marking that cell with the corresponding color.

Each plant can be initialized with a user-defined number of \textit{energy} units. Within the same species, this initial value may vary, for instance due to differences in plant height, age, or total leaf area. In addition, the system is configured to determine the percentage by which a plant’s energy stock increases per discrete simulation step. For each individual plant, the user also specifies a minimum threshold of energy units required for survival. If, during the simulation, a plant’s energy level falls below this threshold, the plant is removed from the grid.

Plants are capable of reproduction over time. The user may specify whether, and after how many simulation steps, an individual plant produces a defined number of offspring of the same species. Each offspring is again initialized with a user-defined number of energy units and is randomly placed in a free grid cell at a distance and direction from the parent plant, both determined within a specified minimum–maximum range. If no suitable free cell is available under these conditions, the offspring is not generated.

Throughout the simulation, each cell occupied by a plant displays the percentage of energy the plant currently possesses relative to its initial energy value. This percentage may exceed or fall below $100\%$. Plants located in immediately adjacent grid cells can form pairwise symbiotic connections, even if they belong to different species. The user initiates such a connection by right-click on one of the involved grid cells. Using a popup window, each corresponding adjacent cell needs to be selected to establish the connection, which is then visualized by a black like. Through these connections, signaling compounds can propagate during the course of the simulation.

\subsection{Configuration of Predators}

The configuration of predators likewise begins with the specification of how many \textit{predator species} are to be included ($0$ to $15$). Predators may act either in swarms or individually. To represent this, we introduce the concept of a \textit{predator cluster}. A predator cluster consists of a freely configurable initial number of individuals (at least one). All individuals within a cluster belong to the same predator species and always occupy the same grid cell simultaneously. Each defined predator cluster can be freely placed by the user on an initial grid cell, regardless of whether the cell is already occupied by a plant and independent of the plant species. Visually, clusters are represented by a dot symbol whose size depends on predator type with the number of individuals indicated either numerically. Multiple predator clusters may occupy the same grid cell simultaneously.

Predator clusters located on grid cells without plants move toward the nearest plant. If several plants are equally close, one is chosen at random. For each predator cluster, the user specifies how many simulation steps are required for the cluster to move into an adjacent grid cell. Predator movement follows an idealized straight line across the grid and is discretized in time by mapping it to the corresponding sequence of grid cells intersected by that line.

\subsection{Configuration of Substances}

As a further element of the system, \textit{substances}, namely \textit{signaling compounds} and \textit{toxins}, are configured. Between $0$ and $15$ distinct substances may be defined. Each substance can be assigned an individual name, and the user must specify whether it is a signaling compound or a toxin.

The configuration of interactions between plants, predator clusters, and substances is carried out using interaction matrices, which the user defines prior to the simulation.

For each signaling compound, it must be specified which plant species are capable of emitting it and which species are receptive to it. Additionally, it must be defined which combinations of predator species trigger the emission, and what minimum number of individuals per predator cluster is required for this trigger. Each signaling compound can be set to propagate either through the air or via grid-cell connections. Airborne spread occurs in concentric circles around the emitting plant. A parameter specifies after how many time steps the radius of this circle expands by one additional grid-cell length. Compounds spreading via grid-cell connections require a parameter indicating after how many time steps adjacent connected cells are reached.

The production of signaling compounds consumes plant energy. Accordingly, each compound is assigned a configurable cost, expressed as the number of energy units deducted from the plant’s energy per time step. A single plant may emit multiple signaling compounds simultaneously. The presence of signaling compounds is visualized by patterns in the affected grid cells. Production ceases once the plant is no longer attacked by the predator species that originally triggered emission. For each compound, the user defines how many additional time steps it remains active after production stops. Once this period expires, the entire effective area originating from the triggering plant is removed, though overlapping active areas generated by other plants of the same compound remain unaffected.

The properties and interactions of toxins are specified in an analogous way. For each toxin, the producing plant species are selected and the energy cost per time step is defined. An interaction matrix determines which combinations of signaling compounds and/or predator species, together with the minimum predator cluster size, trigger toxin production, as well as how many time steps must pass before the toxin takes effect on the plant. Toxins are assumed either to kill predator individuals or to repel predator clusters. The user specifies for each toxin which predator species are targeted and, in the lethal case, how many individuals per cluster are eliminated per time step. In the repellent case, the producing plant is removed from the perception of affected predator clusters, causing them to move toward the next nearest plant without individual losses. The presence of toxins is visualized in the grid cell of the producing plant, using distinct color patterns. Toxin production ceases once the plant is no longer attacked by the triggering predators. For simplicity, toxins are assumed to have no residual effect: once production ends, their influence disappears immediately, leaving the plant both free of the toxin and once again a potential target for predator clusters.

To complete the interaction configuration, the feeding success of predator clusters is specified. For each predator species, the number of energy units derived from consumed plants that is required to generate an additional individual within the cluster is defined. As soon as a predator cluster succeeded in doubling its number of individuals compared to its initial value, the predator cluster becomes separated into two equal independent predator clusters.

\subsection{Configuration of Global Parameters}

Finally, parameters for the simulation run itself are set. The user specifies the duration of a simulation step and chooses among several options for determining the end of the run: either a fixed number of time steps, or one of the following conditions: (1) extinction of a specific plant species, (2) extinction of all plant species, (3) extinction of a specific predator species, (4) extinction of all predator species, (5) reaching a defined threshold for the total plant energy, or (6) reaching a defined threshold for the total number of predator individuals.

\section{Software Implementation}
\label{sec4}

The \textit{Python} programming language, version 3.12.5, was chosen to implement the simulator software. The simulator’s implementation uses as software architecture the \textit{Model–View–Controller} (MVC) pattern. For dependency management, we employed \textit{poetry}\footnote{\url{https://github.com/python-poetry/poetry}}, ensuring the reproducibility of the working environment.

The numerical calculations within the model were carried out using the \textit{NumPy} library \cite{Harris20}. For visualizing the course of a simulation, \textit{Matplotlib} \cite{Hunter07} was employed to generate diagrams.

To prevent simulation results from being lost once the tool is closed, a file management system was implemented. This includes the class \texttt{Export}, responsible for creating files and saving the configured networks, as well as the class \texttt{Import}, which reconstructs stored networks by reloading a selected file back into the tool.

\section{Graphical User Interface and User Manual}
\label{sec5}

The graphical user interface consists of three sections, one of which is a control bar used to define the framework conditions for the grid. This includes specifying the grid size as well as setting the number of different actors and substances.

\begin{figure}[h]
  \centering
  \includegraphics[width=.8\textwidth]{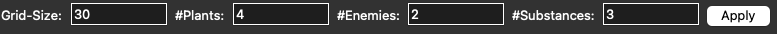}\\[2mm]
  \includegraphics[width=0.6\textwidth]{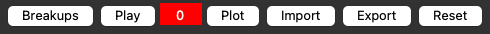}
  \caption{Topbar controls and input fields}
  \label{topbarcontrols}
\end{figure}

In addition, various buttons are available to control the simulation of the configured network and to execute specific actions. Figure \ref{topbarcontrols} (upper part) shows the four input fields, each of which expects a natural number as input. The other controls shown in the lower part are various buttons that are used to perform different actions.

Once valid inputs are provided in the control bar, an empty grid with the corresponding number of actors is generated. Below the control bar, two horizontally aligned sections are displayed: on the left-hand side, input elements for parameter configuration of the actors and substances; on the right-hand side, the visualization of the grid itself.

\begin{figure}[h]
	\centering
	\begin{subfigure}[h]{0.31\textwidth}
        \includegraphics[scale=0.21]{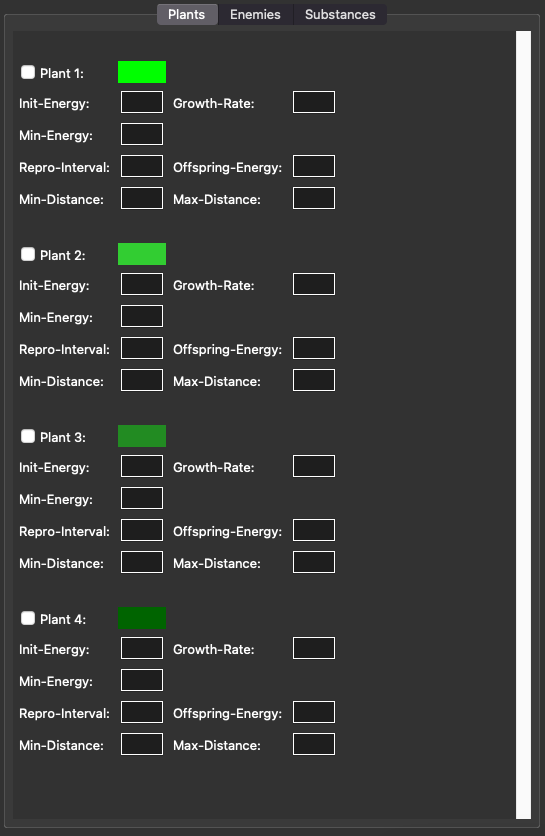}
    \end{subfigure}
    \begin{subfigure}[h]{0.31\textwidth}
        \includegraphics[scale=0.21]{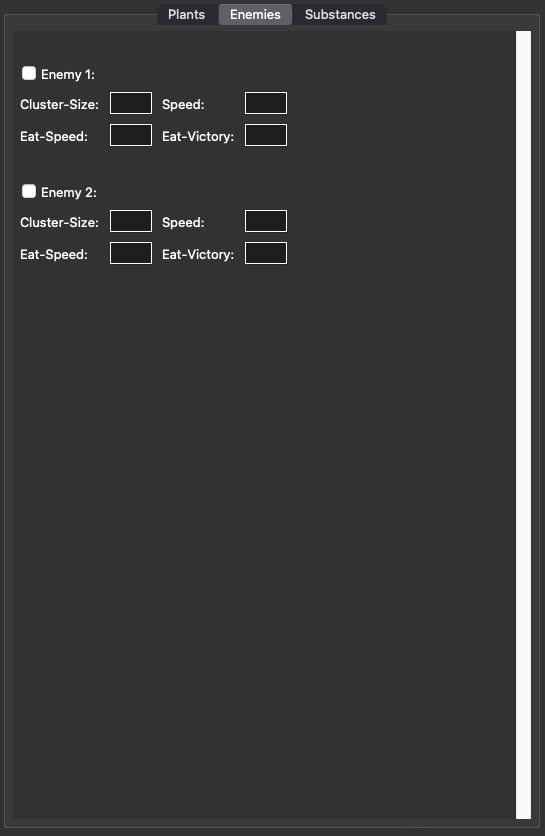}
    \end{subfigure}
    \begin{subfigure}[h]{0.31\textwidth}
        \includegraphics[scale=0.21]{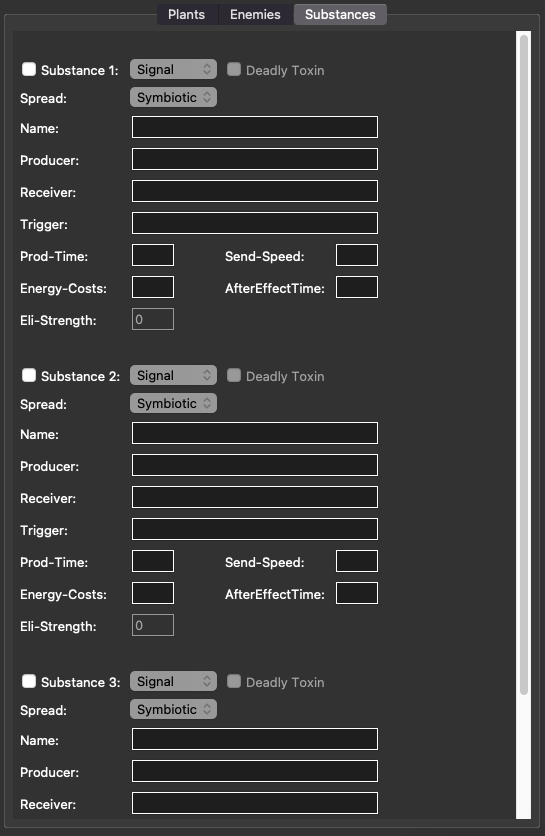}
    \end{subfigure}
    \caption{Tabs for configuration of plants, predators, and substances}
    \label{threetabs}
\end{figure}

The left-hand side for parameter configuration is subdivided into three tabs, see Figure \ref{threetabs}. The first one contains the input fields for the various parameters of the defined plant species followed by a tab that provides the input fields for parameters defined by the user for the predator species. The third tab is reserved for entering the parameters of the different substances.

\subsection{Initialization of the Tool}

When the tool is started, only an empty window with the control bar is displayed. In order to make the parameter tabs and the empty grid visible, the grid size as well as the number of plant species, predator species, and substances must be specified via the input fields and confirmed using the \textit{Apply} button.

However, the input fields for a scenario do not accept arbitrary values but are restricted by predefined limits. They only allow integer values within the model boundaries defined in the implementation. The grid size must be in the range of $1$ to $80$, the number of plant species must be between $1$ and $16$, and the input values for predators and substances may only range from $0$ to $15$. If a value outside these ranges is entered, the tool issues an error message informing the user that the input was invalid and prompts correction of the parameters by displaying the permissible numerical range.

\subsection{Functions of the Control Bar}

Upon successful input, an empty scenario is created in which the parameters for plants, predators, and substances can be individually defined. Actors with completed parameter settings can at this point be placed on the grid.

\begin{figure}[h]
	\centering
	\begin{subfigure}[h]{0.47\textwidth}
		\centering
		\includegraphics[scale=0.4]{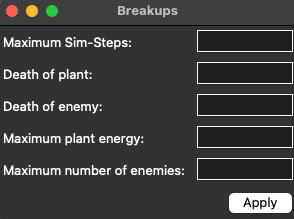}
	\end{subfigure}
	\begin{subfigure}[h]{0.47\textwidth}
		\centering
		\includegraphics[scale=0.4]{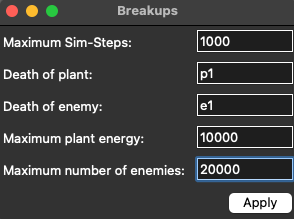}
	\end{subfigure}
	\caption{Defining the conditions for termination of a simulation run}
	\label{conditionsfortermination}
\end{figure}

The \textit{Breakups} button opens a small window in which the different termination conditions for a simulation can be defined. Figure \ref{conditionsfortermination} shows, on the one hand, an empty window when no termination conditions are specified (left), and, on the other hand, a window with completed input fields (right). The latter means that a simulation ends if either $1000$ time steps have elapsed, all plants of species \texttt{p1} have died, all predators of species \texttt{e1} have died, an upper limit of $10,000$ units of plant energy has been reached, or a maximum of $20,000$ predator individuals has been reached.

Next in the control bar is the \textit{Play} button, the function of which is straightforward: it starts the simulation once the user has finished configuring the network and wishes to initiate execution. The right-hand element in the control bar is a color-marked field displaying the current time step of the simulation. Here, a signal color indicates whether the simulation has not yet started, is currently running, or has already ended.

Following this is the \textit{Plot} button. This opens a window in which, for a successfully executed simulation, four different plots representing simulation events can be displayed. To ensure clarity, the plots are organized in tabs between which the user can switch. It is important to note that the plot window is only accessible after a simulation has been executed, since the required information is collected during runtime. Otherwise, the tool blocks access and displays a message instructing the user to first start a simulation before time-series plots can be generated.

The \textit{Import} button opens a window displaying the file manager of the user’s device. There, previously exported files can be located and selected for import into the tool. After a file has been selected, all defined parameters are automatically transferred to the corresponding input fields, and the placement of actors on the grid is restored.

The \textit{Export} button also opens a window, allowing a configured network to be saved as a file under a user-defined name. Since the tool’s file manager is based on the Python module \texttt{pickle}, files are stored in the \texttt{.pkl} format.

The final button in the control bar is the \textit{Reset} button. It restores a completed simulation to its initial state, enabling a new simulation run without requiring the network to be reconfigured or re-imported.

\subsection{Preparing the Grid}

In order to place plants on the grid, the corresponding input fields in the tab for the plant species to be placed must first be completed. A plant species is defined by seven different input fields, which are listed in Table \ref{parametersofaplantspecies}.

\begin{table}
\begin{tabular}{|l|l|l|}
\hline
\textbf{Parameter} & \textbf{Description} & \textbf{Range}\\
\hline
\hline
Init-Energy & Initial energy of a plant & $\mathbb{N}$\\
\hline
Growth-Rate & Growth rate in $\%$ of the initial energy & $\mathbb{N}$\\
\hline
Min-Energy & Minimum energy required for survival & $\mathbb{N}$\\
\hline
Repro-Interval & Interval after which offspring are produced & $\mathbb{N}\setminus\{0\}$\\
\hline
Offspring-Energy & Initial energy of the offspring & $\mathbb{N}$\\
\hline
Min-Distance & Minimum dispersal distance for offspring & $\mathbb{N}$\\
\hline
Max-Distance & Maximum dispersal distance for offspring & $\mathbb{N}$\\
\hline
\end{tabular}
\caption{Parameters of a plant species}
\label{parametersofaplantspecies}
\end{table}

It is important to note that the input field for the reproduction interval may alternatively be left blank instead of entering a $0$ if the plant is not intended to reproduce. In this case, the graphical interface interprets the empty field as $0$.

Furthermore, for the offspring input fields, it is taken into account that if the maximum dispersal distance is accidentally set to a smaller value than the minimum distance, the maximum distance is automatically adjusted to the value of the minimum distance plus 1. This ensures that the maximum distance is always greater than the minimum distance.

A predator species, defined in the predator tab, is characterized by four input fields for its parameters, which are listed in Table \ref{parametersofapredatorspecies}.

\begin{table}
\begin{tabular}{|l|l|l|}
\hline
\textbf{Parameter} & \textbf{Description} & \textbf{Range}\\
\hline
\hline
Cluster-Size & Number of individuals in the cluster & $\mathbb{N}\setminus\{0\}$\\
\hline
Speed & Movement speed in time steps & $\mathbb{N}$\\
\hline
Eat-Speed & Eating rate in units of plant energy & $\mathbb{N}$\\
\hline
Eat-Victory & Energy required to generate one offspring & $\mathbb{N}\setminus\{0\}$\\
\hline
\end{tabular}
\caption{Parameters of a Predator Species}
\label{parametersofapredatorspecies}
\end{table}

For signaling compounds, there are more options and input fields due to the different types. First, the user can decide via a selection menu whether the substance is a signaling compound or a toxin. If it is a signaling compound, the user can then specify whether it propagates through the air or via grid connections.

If it is a toxin, however, the propagation menu is blocked so that no selection can be made. Instead, the checkbox to the right of the substance type becomes active, allowing the user to define whether the toxin has a lethal effect. If the box is not checked, the toxin exerts a repelling effect on predators.

Each substance can be assigned an arbitrary name. This input is not subject to any specific requirements and may contain both letters and numbers, entirely according to the user’s preference.

The second input field specifies the plant species capable of producing the substance. A plant species is entered as \texttt{p} followed by a number between \texttt{1} and \texttt{16} corresponding to the species. If multiple plant species produce the same substance, several entries can be provided, separated by commas. The same principle applies to the next input field, in which the plant species that can receive a signaling compound are defined.

The next input field, which defines the triggers of a substance, offers different options depending on whether the substance is a signaling compound or a toxin.

For a signaling compound, the predator species that activate the signal are specified by entering `\texttt{e}` followed by a number between \texttt{1} and \texttt{15}. Since cluster size also plays a role, it is entered after the predator species, separated by a comma. This pair of inputs forms a trigger condition that must be met for the signaling compound to be released. If multiple triggers are required, they are separated by semicolons. An accepted input could, for example, be: \texttt{e1,1; e2,5; e3,2}

For toxins, a trigger consists of three elements, each separated by a comma. In addition to the trigger elements already used for signaling compounds, the name of the signaling compound must also be specified, and it must be placed at the beginning of the trigger. In other words, toxin triggers are defined by first the signaling compound name, then the predator species, and finally the cluster size. As with signaling compounds, multiple triggers are separated by semicolons. An accepted input could, for instance, be: \texttt{s1,e1,1; s2,e2,5; s1,e3,2}. Table \ref{inputfieldsfornumericalsubstanceparameters} gives an overview about input fields for numerical substance parameters.

\begin{table}
\begin{tabular}{|l|l|c|c|}
\hline
\textbf{Parameter} & \textbf{Description} & \textbf{Signal Substance} & \textbf{Toxin} \\
\hline
\hline
Prod-Time & Production time in time steps & Yes & Yes \\
\hline
Energy-Costs & Energy costs per time step & Yes & Yes \\
\hline
Eli-Strength & Mortality rate per time step & No & Yes \\
\hline
Send-Speed & Transmission speed in time steps & Yes & No \\
\hline
AfterEffectTime & Duration of aftereffect in time steps & Yes & No \\
\hline
\end{tabular}
\caption{Input fields for numerical substance parameters.}
\label{inputfieldsfornumericalsubstanceparameters}
\end{table}

At this stage, all essential components required for placing the agents on the grid have been addressed.
In order to position plants or predator clusters on the grid, the corresponding checkbox must be activated. Activating the checkbox selects the agent, thereby enabling its placement on the grid.
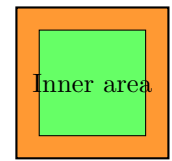
\begin{floatingfigure}[r]{0.25\linewidth}
\begin{tikzpicture}
\draw[thick,fill=orange!80] (0,0) rectangle (2,2);
\draw[fill=green!60] (0.3,0.3) rectangle (1.7,1.7);
\node at (1,-0.3) {Outer area};
\node at (1,1) {Inner area};
\end{tikzpicture}
\caption{Inner and outer area of a grid cell}
\label{innerandouter}
\end{floatingfigure}

For substances, the procedure differs, since they are coupled to plants. Substances also provide a checkbox, which must be activated to inform the tool which substances are to be included in the simulation.

Each grid cell consists of two areas, an inner and an outer section, which can be represented visually as in Figure \ref{innerandouter}.

The inner area is responsible for displaying the colors of the selected plants, which fill this section once placed. The outer area, by contrast, indicates the availability of a substance, which is represented in different colors depending on the substance type. The specific colors and their corresponding meanings can be found in Table \ref{Meaningofcolorsforsubstanceactivity}.

\begin{table}
\centering
\rowcolors{2}{gray!10}{white}
\begin{tabular}{|>{\columncolor{white!60}}m{2cm}|m{10cm}|}
\hline
\textbf{Color} & \textbf{Meaning} \\
\hline
\hline
\rowcolor{yellow!60}
Yellow & At least one signaling substance is being produced and none is active \\
\hline
\rowcolor{orange!80}
Dark Orange & At least one signaling substance is active \\
\hline
\rowcolor{orange!50!red!40}
Coral & At least one toxic substance is being produced and none is active \\
\hline
\rowcolor{red!70}
Red & At least one toxic substance is active \\
\hline
\end{tabular}
\caption{Meaning of colors for substance activity.}
\label{Meaningofcolorsforsubstanceactivity}
\end{table}

For signaling substances, the radius is visualized by means of an orange highlight applied to the respective fields, with the outer area being filled to indicate to the user where a signaling substance is active. Since multiple signaling substances can be active on the same grid cell when spreading through the air, the color is rendered progressively darker for each additional substance.

For signaling substances transmitted via grid connections, the connection must be manually established between plants. This is accomplished by right-clicking on a plant to open a popup window displaying a list of neighboring plants. A connection to a neighboring plant can then be established by activating the corresponding checkbox.

A grid connection is displayed as a black line between neighboring plants. However, once a signaling substance is transmitted through the connection, the color of the line changes to gray.

Predator clusters are represented by small markers on the grid cell, which can appear in two different colors. They are shown in dark blue when the predators are not lethally poisoned, i.e., when they are in their normal state. If they become lethally poisoned, the marker on the grid is displayed in red.

Within the graphical interface, two mechanisms exist for increasing the size of a cluster. The first method is based on classical growth, in which the number of individuals within a cluster increases through the intake of plant energy. However, this approach results in static simulation behavior. Because predator reproduction relies solely on this principle, it is difficult to generate a dynamic interplay between plants and predators in which dominance alternates over the course of the simulation.

To increase the dynamism of the simulation, an additional logic has been implemented. This logic checks whether the number of individuals has doubled relative to the initial cluster size. Once this condition is met, the cluster is divided into two separate clusters. In the graphical interface, this is visualized by the appearance of two markers, each representing half of the individuals of the original cluster.

To prevent the number of clusters from growing without limit, the maximum number of clusters is restricted to half of the available grid cells. This constraint enables predators to reach a certain size, while still allowing plants a chance to persist and potentially displace the predators again. The exact course of the simulation, however, depends primarily on the chosen parameters.

\begin{figure}[h]
\centering
\includegraphics[width=0.9\textwidth]{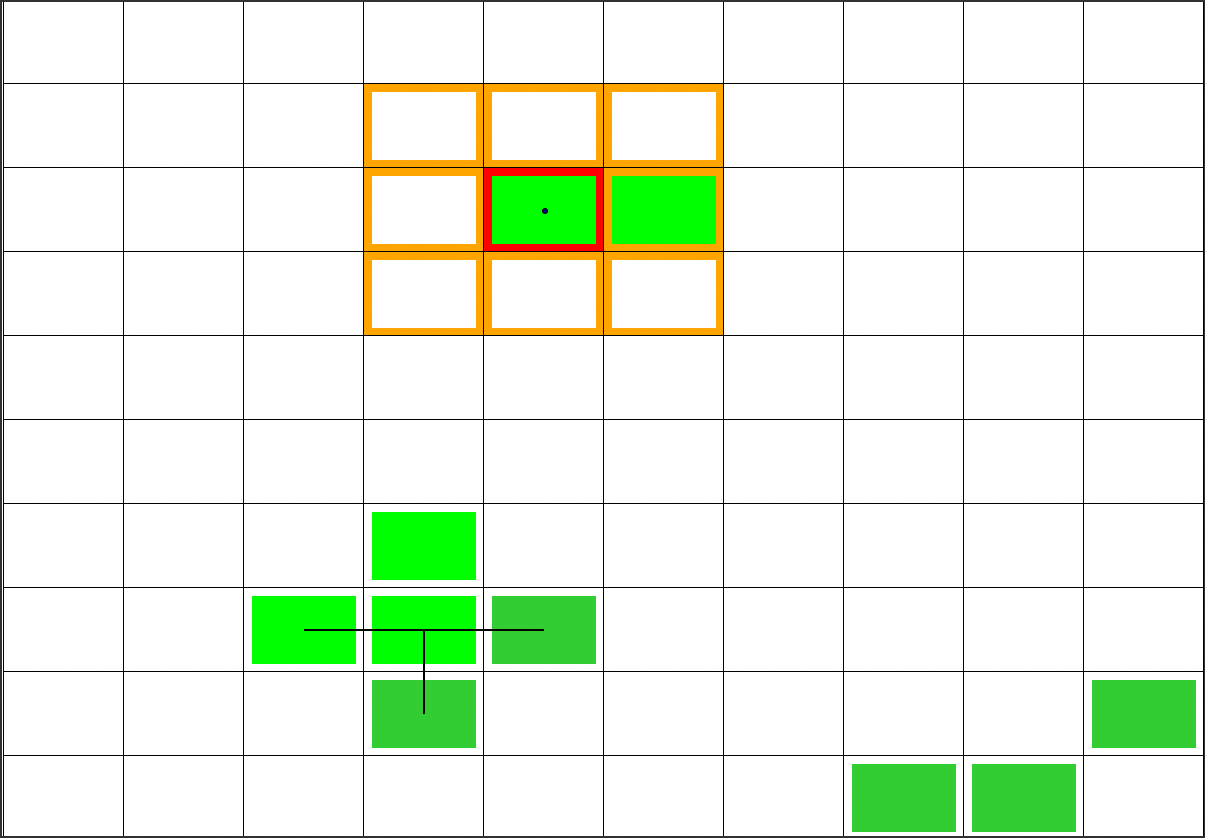}
\caption{Example of a predefined grid with plants (dark and light green) and predators (black dot). A network of roots connecting neighbored plants is represented by black lines across adjacent cells. Grid cells affected by signaling substances have been marked with a dark orange outer area while presence of a toxin is framed in red.}
\label{predefinedgridwithplants}
\end{figure}

Figure \ref{predefinedgridwithplants} illustrates how the grid with plants and predators can appear during a simulation.

\section{Case Study}
\label{sec6}

We conducted a simulation case study to demonstrate the functionality of the tool and to visualize the results using a number of plots. The simulation extends over a longer time horizon, allowing temporal changes to be represented and analyzed in the plots. First, a scenario is configured that can be entered into the tool. The simulation results are then presented separately for plants and their predators.

\subsection{Configuration of the Network}

To ensure a meaningful dynamic within the simulation, a biotope with a size of $15 \times 15 = 225$ grid cells is selected. Within this biotope, two different plant species, two predator species, and three different substances are present. The agent configuration is based on specific parameters, which are summarized in Table \ref{selected-parameters-of-the-agents-for-the-simulation-study}.

\begin{table}
\centering
\begin{tabular}{|l|c|c|l|c|c|}
\hline
\multicolumn{3}{|c|}{\textbf{Plants}} & \multicolumn{3}{c|}{\textbf{Predators}} \\
\hline
\textbf{Parameter} & \textbf{Plant 1} & \textbf{Plant 2} & \textbf{Parameter} & \textbf{Enemy 1} & \textbf{Enemy 2} \\
\hline
Init-Energy & 123 & 90 & Cluster-Size & 30 & 10 \\
\hline
Growth-Rate & 3 & 2 & Speed & 1 & 4 \\
\hline
Min-Energy & 20 & 10 & Eat-Speed & 1 & 10 \\
\hline
Repro-Interval & 30 & 35 & Eat-Victory & 20 & 30 \\
\hline
Offspring-Energy & 30 & 50 & & & \\
\hline
Min-Distance & 1 & 2 & & & \\
\hline
Max-Distance & 2 & 6 & & & \\
\hline
\end{tabular}
\caption{Selected parameters of the agents for the simulation study.}
\label{selected-parameters-of-the-agents-for-the-simulation-study}
\end{table}

The most significant difference between the two plant species lies in their mode of reproduction: the first species reproduces exclusively locally, whereas the second species disperses its offspring within a larger radius. To balance this difference, the first plant is assigned a shorter reproduction interval than the second.

The primary difference between the two predator species concerns the energy requirements for reproduction and the energy intake per time step. While the predators of the first species require $20$ time steps to produce a new individual, the predators of the second species demand more energy for reproduction but simultaneously consume more energy per time step. The parameters selected for the substances in the simulation are listed in Table \ref{selected-parameters-of-the-substances-for-the-simulation-study}.

\begin{table}
\centering
\begin{tabular}{|l|c|c|c|}
\hline
\textbf{Parameter} & \textbf{Substance 1} & \textbf{Substance 2} & \textbf{Substance 3} \\
\hline
Type             & Signal & Toxin & Toxin \\
Deadly           & --     & No    & Yes   \\
Spread           & Air    & --    & --    \\
Name             & \texttt{s1}     & \texttt{s2}    & \texttt{s3}    \\
Production       & \texttt{p1,p2}  & \texttt{p1}    & \texttt{p2}    \\
Receiver         & \texttt{p1,p2}  & --    & --    \\
Trigger          & \texttt{e1,10; e2,30} & \texttt{s1,e1,20; s1,e2,40} & \texttt{s1,e1,40; s1,e2,15} \\
Prod-Time        & 3      & 3     & 2     \\
Send-Speed       & 10     & --    & --    \\
Energy-Costs     & 1      & 1     & 1     \\
AfterEffectTime  & 1      & --    & --    \\
Eli-Strength     & --     & --    & 6     \\
\hline
\end{tabular}
\caption{Selected parameters of the substances for the simulation study.}
\label{selected-parameters-of-the-substances-for-the-simulation-study}
\end{table}

\begin{figure}[t]
	\centering
	\includegraphics[scale=0.25]{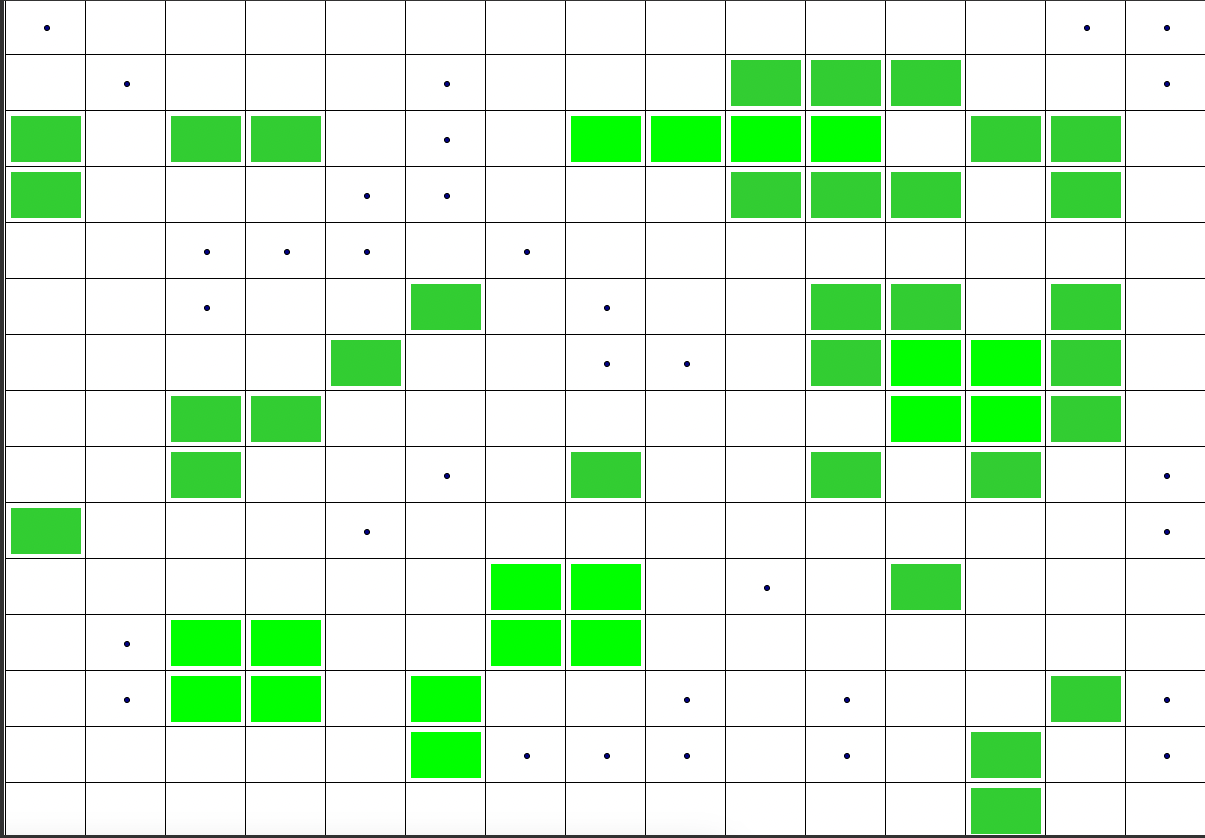}
	\caption{Initial population of the grid for the simulation case study}
    \label{sim-grid}
\end{figure}

The plant of the first species pursues a repelling strategy, whereas the plants of the second species employ a lethal strategy. To provide a warning to other plants, a signal substance is dispersed through the air, with its effective radius increasing every 10 time steps. For the initial population of the grid, a diverse mixture of the two plant species and  the two predator species was selected in order to simulate a realistic biotope, see Figure \ref{sim-grid}. The plants were arranged in different constellations: in some cases adjacent to individuals of the same species, and in other cases adjacent to individuals of the other plant species. A similar approach was applied to the predators, which were evenly distributed across the grid.  

\subsection{Simulation Results}

Here, the results of the simulation study are presented in the form of plots. Since the simulation was conducted with two plant species and two predator species, each plot contains three time-series curves. The black dashed line represents the sum of the respective measured values.

\subsubsection*{Development of the Plant Population over Time}

\begin{figure}[h]
	\centering
	\includegraphics[scale=0.49]{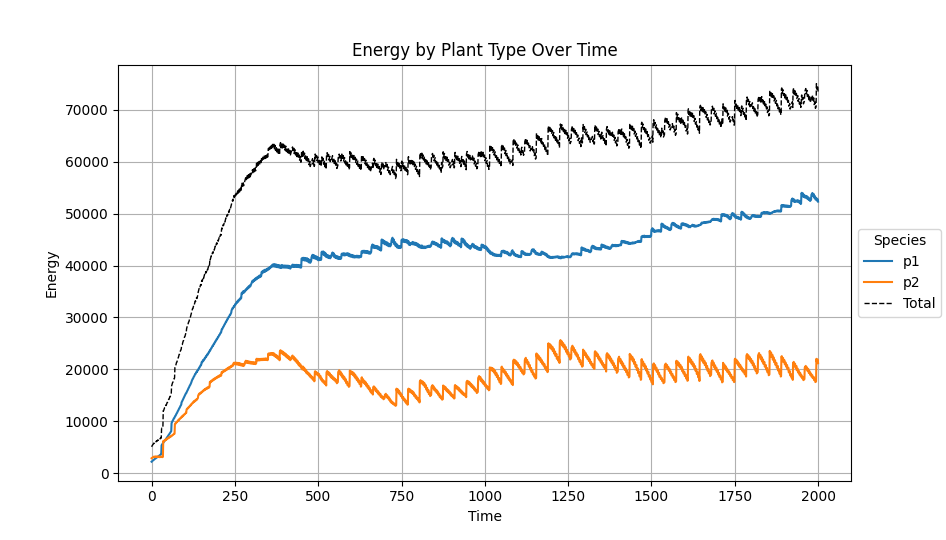}
	\caption{Time course of the energy of each plant species}
	\label{sim-plot1}
\end{figure}

For the plant population, the energy of each plant species was recorded in order to represent it as an aggregated time-series curve for each species. To achieve this, all plants of the respective species were iterated over, and for each time step the currently available energy was summed. The aggregated values are visualized in Plot \ref{sim-plot1} as a bundled line.

\begin{figure}[h]
	\centering
	\includegraphics[scale=0.49]{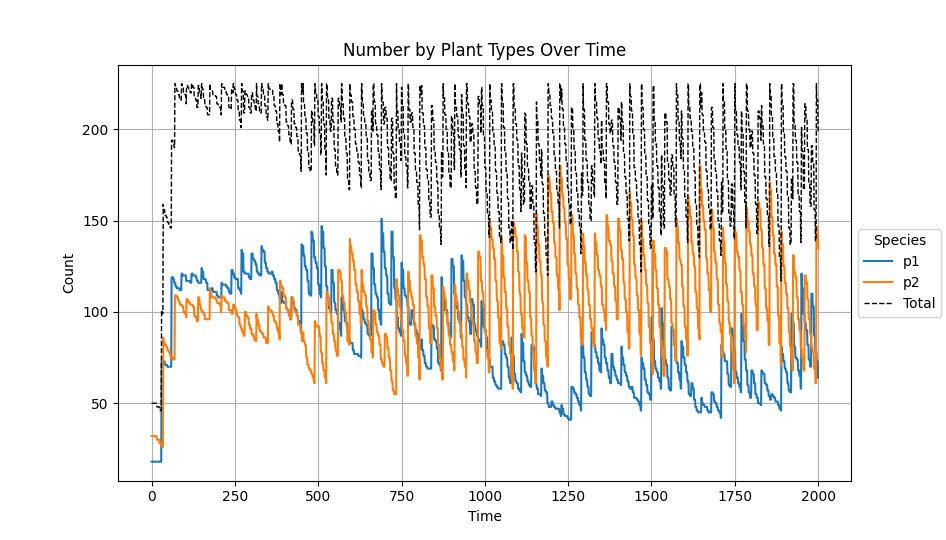}
	\caption{Time course of the occurrence of each plant species}
	\label{sim-plot2}
\end{figure}

The number of plants of each species was determined at every point in time, see Plot \ref{sim-plot2}. It shows, on the one hand, how many plants of a given species are present on the grid at each time step. On the other hand, the total number of all plants on the grid is also represented, shown as the black line.

\subsubsection*{Development of the Predator Population over Time}

\begin{figure}[h]
	\centering
	\includegraphics[scale=0.49]{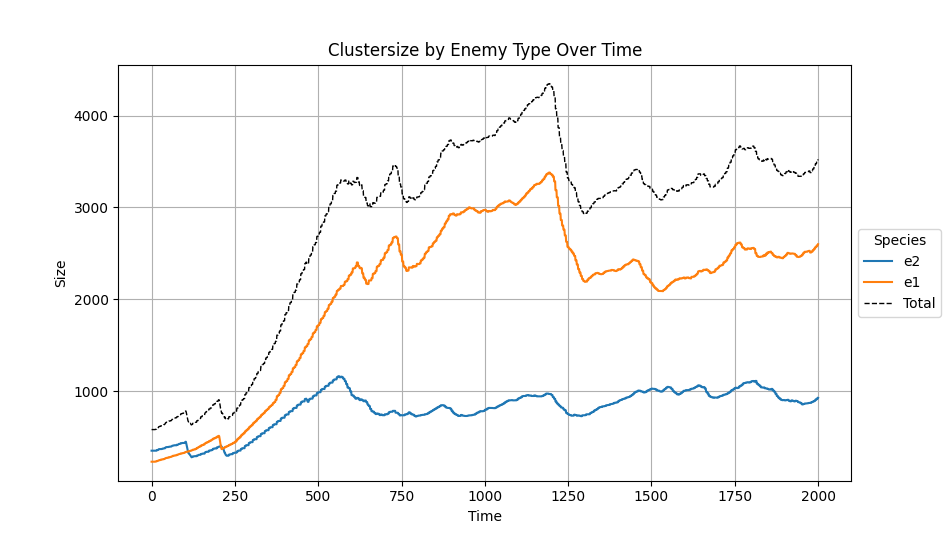}
	\caption{Time course of the individuals of each predator species}
	\label{sim-plot3}
\end{figure}

\begin{figure}[h]
	\centering
	\includegraphics[scale=0.49]{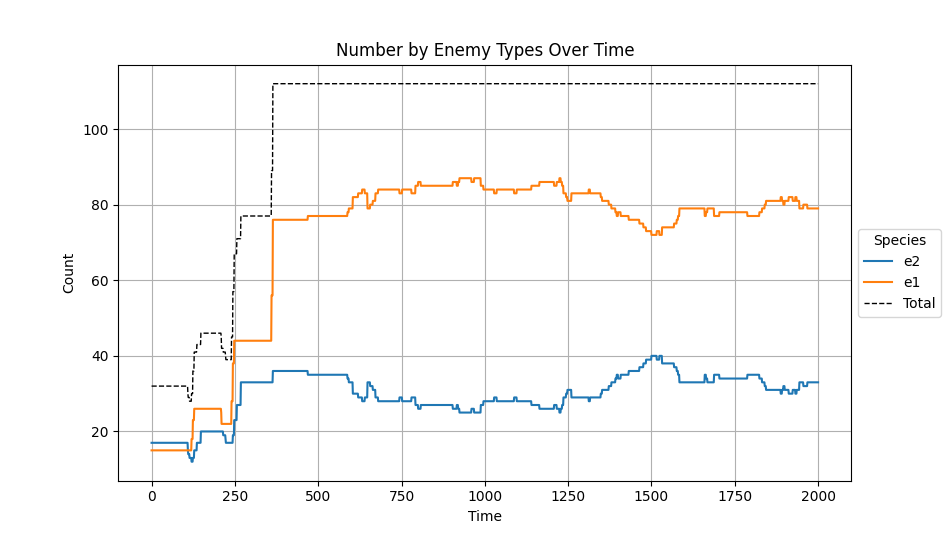}
	\caption{Time course of the number of predator clusters for each predator species}
	\label{sim-plot4}
\end{figure}

Plot \ref{sim-plot3} exhibits the number of individuals of a specific species across all placed predator clusters over the course of time. Similar to the plants, Plot \ref{sim-plot4} shows the number of clusters of the different predator species over time.

The results clearly demonstrate the dynamic changes in the populations of plants and predators over time.  
To obtain meaningful insights, the simulation should run for at least $1000$ steps, as all four plots exhibit an initial transient phase during which the model stabilizes.  
The duration of this phase depends on the initial configuration.  
Longer simulation periods generally lead to more distinct results. For better comparability with the results presented in Figures \ref{sim-plot1} to \ref{sim-plot4}, the following section shows the outcomes of a simulation with $4000$ time steps under identical initial conditions.

\begin{figure}[h]
	\centering
	\begin{subfigure}[h]{0.49\textwidth}
		\centering
		\includegraphics[width=\textwidth]{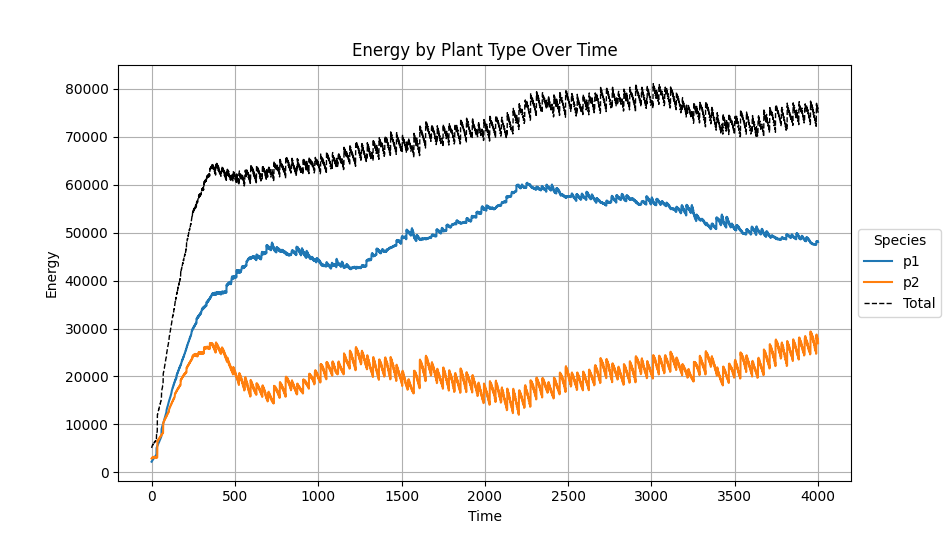}
		\caption{Energy of plants over time}
		\label{subfig:sim-plot1-4000}
	\end{subfigure}
	\begin{subfigure}[h]{0.49\textwidth}
		\centering
		\includegraphics[width=\textwidth]{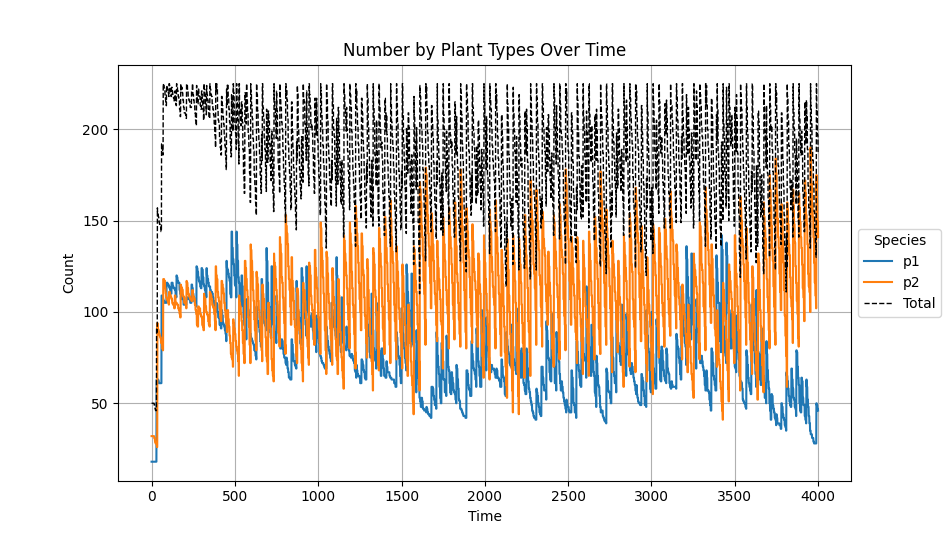}
		\caption{Number of plants over time}
		\label{sim-plot2-4000}
	\end{subfigure}

	\begin{subfigure}[h]{0.49\textwidth}
		\centering
		\includegraphics[width=\textwidth]{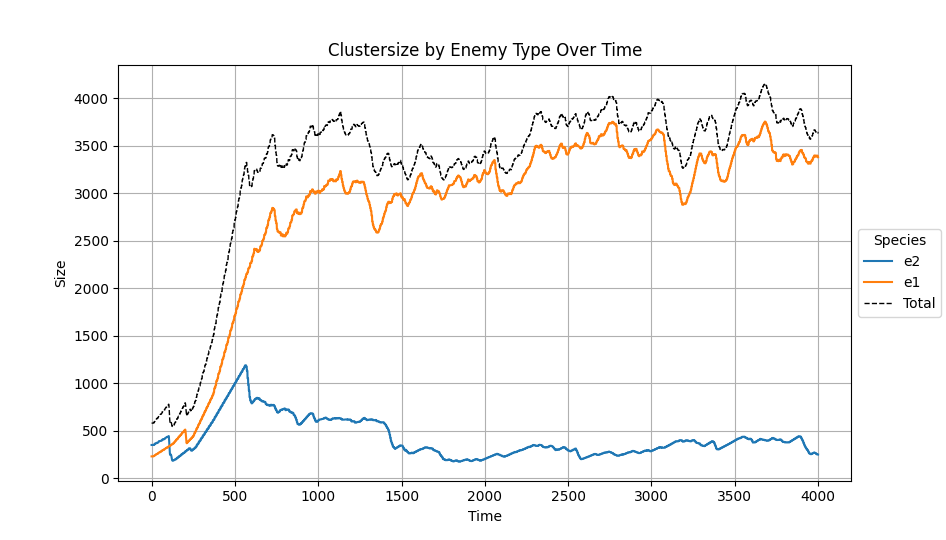}
		\caption{Individuals of predators over time}
		\label{subfig:sim-plot3-4000}
	\end{subfigure}
	\begin{subfigure}[h]{0.49\textwidth}
		\centering
		\includegraphics[width=\textwidth]{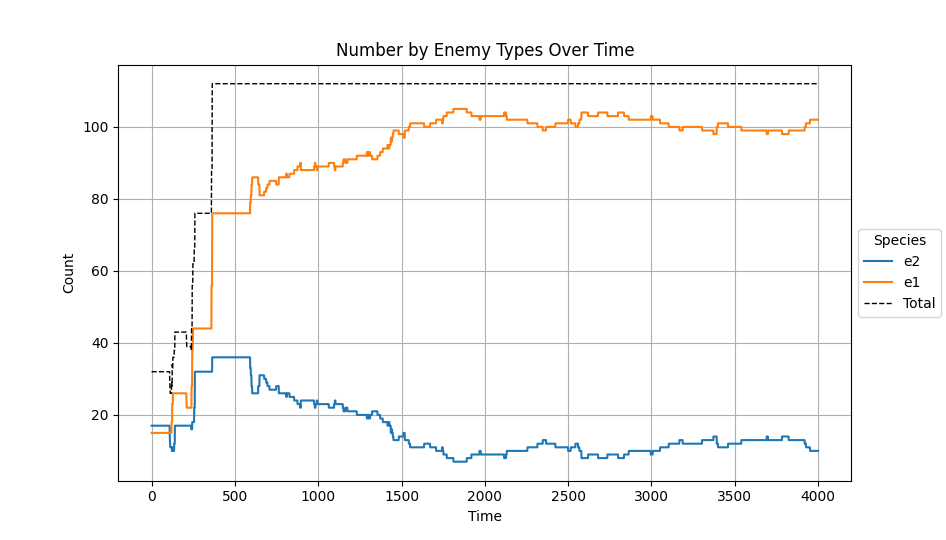}
		\caption{Number of predator clusters over time}
	\end{subfigure}
	\caption{Simulation over $4000$ time steps}
    \label{sim-plots_4000}
\end{figure}

In these results, the dynamics of system development are clearly discernible, particularly in Plot \ref{sim-plots_4000}a, which depicts the energy of the plants. A wave-like pattern emerges, indicating that the availability of energy for the plants is subject to temporal fluctuations.

The results concerning the number of plants of a specific species also exhibit dynamic behavior in the form of wave-like trajectories. This dynamic pattern is even more evident here than in the results regarding the energy levels of the plants, where it is also present. Furthermore, in cases where multiple plant species are involved, in this case two, an inverse dynamic can be observed. This means that an increase in the population of one species corresponds to a decrease in the population of the other species, and vice versa.

The plots showing the number of individuals of a specific predator species likewise reveal an initial adaptation phase, during which the population level adjusts to the given conditions. Subsequently, the number of individuals oscillates in an irregular dynamic around an approximately constant level. These fluctuations depend, among other factors, on which plants the predators preferentially consume and how the habitat is populated with plants. In addition, an element of randomness plays a role, as the offspring of plants are placed randomly and their number varies.

A closer inspection of Plots \ref{sim-plots_4000}b and \ref{sim-plots_4000}c shows that at around time step $1300$, a local minimum in the number of individuals of predator species \texttt{e1} occurs, while at the same time the number of plants of species \texttt{p2} reaches a local maximum.

The results discussed thus far display behavior that can also be observed in the Lotka–Volterra model. This model consists of two differential equations and describes the interactions between predator and prey within a system \cite{anisiu14}. Its corresponding curves reflect the population sizes within the system over time.

The generated plot of the number of clusters with different predator species shows that, overall, the maximum possible number of clusters on the grid remains nearly constant. This can be explained, on the one hand, by the choice of the maximum cluster number, which can emerge during the simulation through division, and on the other hand by the selected parameter values.

At the same time, a slightly periodic behavior can be observed in the number of individual clusters of a specific species, similar to the patterns seen in other results. This behavior indicates that the number of clusters of a given species increases and decreases over time. Once again, an inverse dynamic between the clusters of different species can be observed.

Another noteworthy aspect is the fact that the progression of a simulation depends heavily on the chosen parameters. It has been shown that even small modifications to a single parameter can lead to significant changes in the simulation outcomes.

For better comparability, the exact same configuration of the system has been employed with respect to the number of actors, substances, and population of the grid. In short, the configured system was imported into the tool and subjected to a minor adjustment concerning a single value in the input for the trigger of the lethal substance, which is now activated at $35$ individuals of predator species \texttt{e2}, instead of $15$ as in the simulation run before.

\begin{figure}[h]
	\centering
	\begin{subfigure}[h]{0.49\textwidth}
		\centering
		\includegraphics[width=\textwidth]{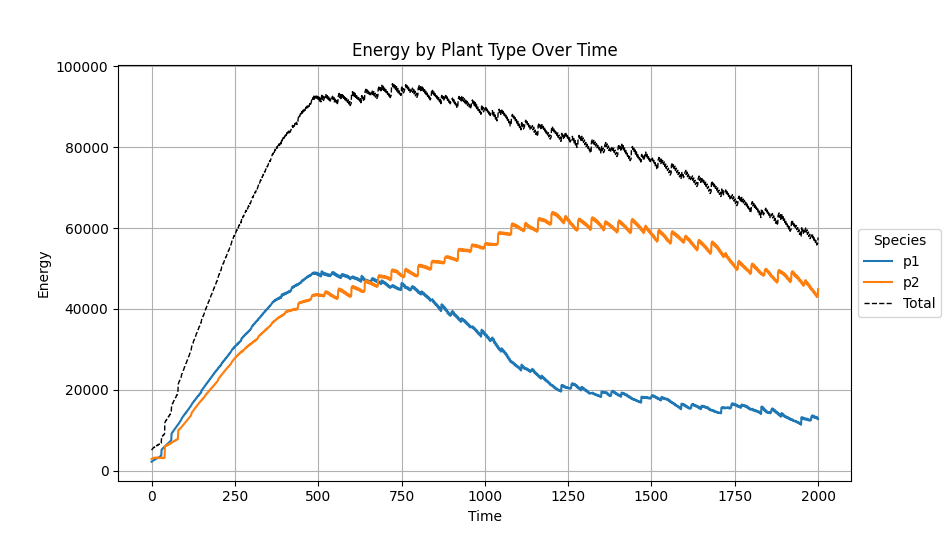}
		\caption{Plant energy over time}
		\label{subfig:sim-plot1-2000_35}
	\end{subfigure}
	\begin{subfigure}[h]{0.49\textwidth}
		\centering
		\includegraphics[width=\textwidth]{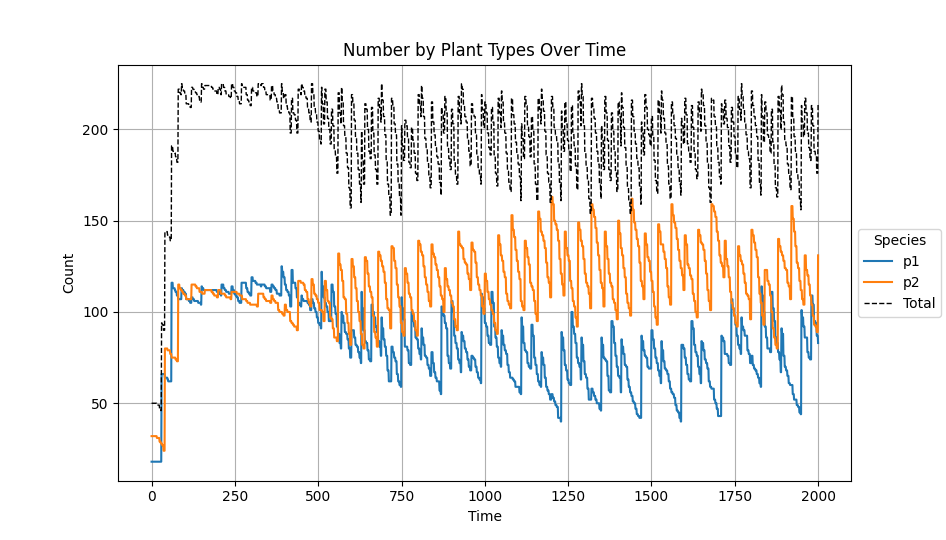}
		\caption{Number of plants over time}
		\label{subfig:sim-plot2-2000_35}
	\end{subfigure}
	
	\begin{subfigure}[h]{0.49\textwidth}
		\centering
		\includegraphics[width=\textwidth]{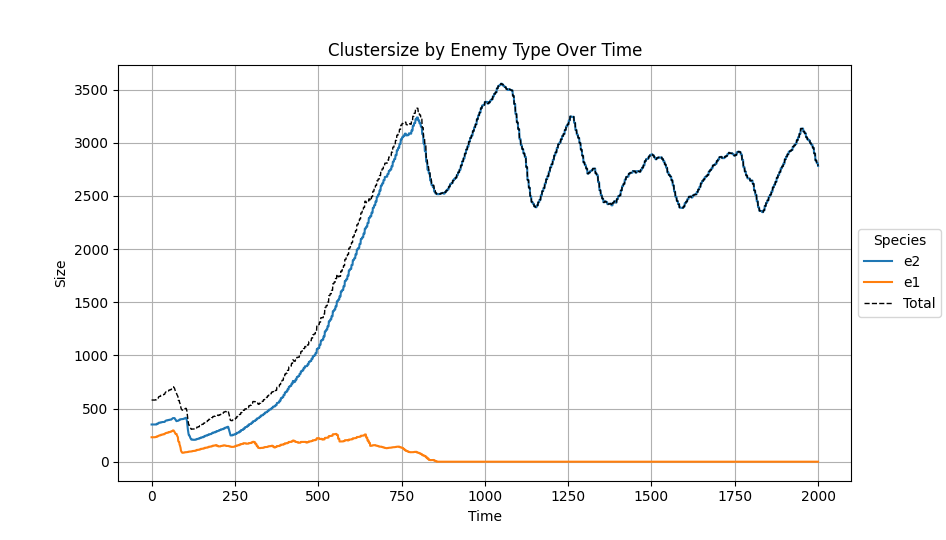}
		\caption{Enemy individuals over time}
		\label{subfig:sim-plot3-2000_35}
	\end{subfigure}
	\begin{subfigure}[h]{0.49\textwidth}
		\centering
		\includegraphics[width=\textwidth]{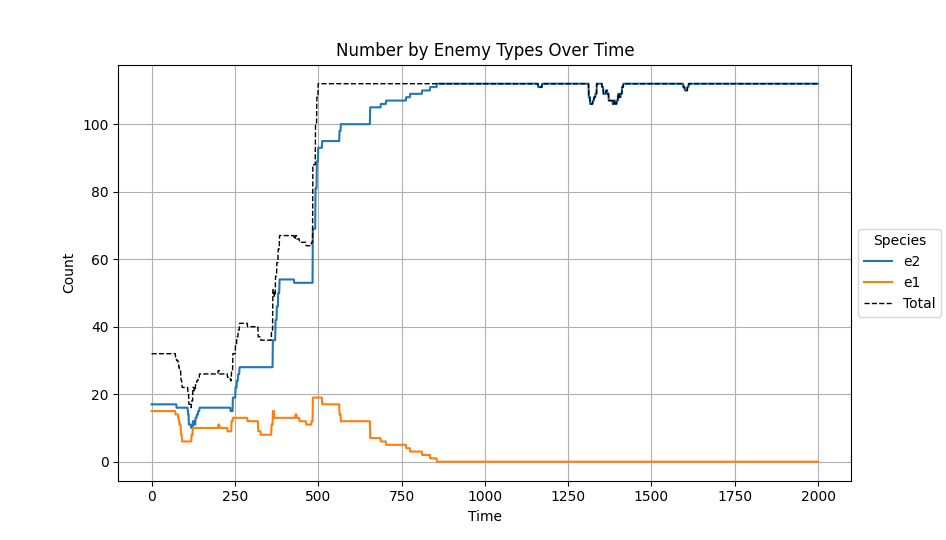}
		\caption{Number of predator clusters over time}
		\label{subfig:sim-plot4-2000_35}
	\end{subfigure}
	\caption{Simulation over $2000$ time steps with merely one slight parameter modification: The lethal toxin is now activated at a higher threshold of $35$ individuals from predator species \texttt{e2} instead of $15$.}
    \label{sim-plots-2000_35}
\end{figure}

The effects of the small change are clearly visible. By slightly increasing the threshold of individuals required for predator type \texttt{e2}, predator type \texttt{e1} can no longer sustain itself over the long term and dies out relatively quickly. At the same time, predator type \texttt{e2} becomes established and exhibits periodic behavior, as can be observed in Plot \ref{sim-plots-2000_35}c.

It can be stated that the simulation results strongly depend on the chosen parameters. Even small modifications to the parameters can lead to significantly different outcomes, despite the initial conditions being nearly identical. In addition to the initial parameters, randomness also plays a crucial role in the course of the simulation. The predators’ decisions regarding which plant to target next may have far-reaching consequences and, in the worst case, lead to the extinction of a predator cluster.

During the simulation run, the grid including hosted plants and predator clusters undergoes a successive progression over time that is visualized on the fly along with each dicrete time step. In this way, dynamical effects of processes become visible, and correlations between reasons and resulting outcome can be understood.

The configuration of predator-pray systems that can persist for a \textit{long time} turns out to be a demanding but productive task. A subsequent analysis of those systems might reveal the essential prerequisites for a setting of plants that is robust and capable of regeneration in an appropriate amount of time. Spatial distributions of plants in a landscape result from upstream evolutionary and behavioral optimization caused by accumulated experiences with numerous predators over time. Our simulation software enables an \textit{in-silico} experimental verification of successful and robust spatial arrangements of plants in conjunction with a well-balanced mixture of signaling substances and toxins.

In the scenario addressed in the case study, two types of plant species have been taken into account. On the one hand, slowly growing plants able to produce highly effective toxins and on the other hand, rapidly growing plants with limited capacities in production of toxins. Both types of plants interact symbiotically, as slowly growing plants have been closely surrounded by rapidly growing plants. In this combination, herbivores need to invest much more energy and effort to sustainably damage both types of plants. Moreover, fast rehabilitation after a predator attack seems feasible. \textit{Insulation spots} spread over the landscape with a specific choice of plant types forming a spot could be the clue to combine effective defense with low energy costs. This principle extends protection by simple neighborhood watch.

\section{Discussion and Conclusions}
\label{sec7}

In terms of natural computation, we adopt most relevant behavioral patterns of plants for protection against herbivores. These patterns include spatial distribution of plants on the ground, molecular signaling mechanism over air or through a symbiotic network of roots when attacks occur, and the production of highly effective toxins. They act in a way either to kill predators or to repel them. Our simulation software \textit{PlantProtectionSim} enables a detailed and flexible grid configuration together with initial assignments with plants and predators of different types. Furthermore, substances might be specified to form signaling molecules, bitter compounds, or toxins. Due to the high degree of configurability, the tool can capture a large variety of scenarios to be considered as topological optimization processes. All underlying elementary events, like reproduction of a single plant or spreading of signaling molecules for the duration of a time step, form elementary operations on a moderate level of abstraction. The energy of plants and the number of individuals in predator clusters serve as fitness measures to indicate survival and extinction. The visual representation of each simulation run along with the generation of diagrams and charts for analysis illustrates the dynamics of the system under study.

Our simulation software intends to conduct numerous simulation case studies with varying configuration scenarios taken from real life or from desired outcome to find corresponding initial data to produce and reproduce a specific result. The behavioral patterns with all events and actors can be seen as a form of computation when progressing in time and space. Technical applications for construction of secure wireless sensor networks have been envisioned.

From a more general point of view, our approach based on a fine-grained grid with progression of successive discrete events provides a flexible simulation environment for a plethora of case studies beyond those covered by differential equations from the conventional Lotka-Volterra form. We strive for identification of critical parameters and corresponding values crucial for the behavioral pattern as a whole. Especially tipping points and their approximation are of special interest in order to understand what makes the difference between survival and extinction over the long term.

Further work will take into consideration the integration of wind and its direction since distribution of molecules in the air is influenced by airstream, and some herbivores did adapt by feeding contrarily. Moreover, we plan an extension towards an artificial evolution able to achieve optimal grid topologies and settings from scratch with high robustness against specific forms of attacks.

\bibliography{sn-bibliography}

\end{document}